# Inhibiting phase drift in multi-atom clocks using the quantum Zeno effect

S.U. Shringarpure* and J.D. Franson

Physics Department, University of Maryland Baltimore County, Baltimore, MD 21250 USA
*saurabh.s@umbc.edu

## ABSTRACT

The accuracy of an atomic clock depends in part on the bandwidth of the relevant atomic transitions. Here we consider an ensemble of $N$ atoms whose transition frequencies have been independently perturbed by environmental effects or other factors. We consider the possibility of using the quantum Zeno effect to lock the relative phase of the atoms, which would decrease their effective bandwidth by a factor of $1/\sqrt{N}$. We analyze an example in which the quantum Zeno effect can be used to lock the relative phase of a pair of atoms, after which the elapsed time can be determined. Practical applications may require $N \gg 1$ in order to achieve a good signal-to-noise ratio.

## Introduction

    Atomic clocks have a number of important applications[1-5], and there has been considerable progress in developing new techniques to improve their performance[6-9]. The precision of atomic clocks depends in part on the absorption bandwidth of the relevant atomic transition[10]. Here we consider an ensemble of $N$ atoms whose transition frequencies have been independently perturbed by a small amount due to coupling to the environment or other factors, such as the effective bandwidth due to finite lifetimes. We investigate the possibility of using the quantum Zeno effect to lock the relative phases of the atoms, which would decrease their effective bandwidth by a factor of $1/\sqrt{N}$. An example is analyzed in which the quantum Zeno effect can be used to lock the relative phase of a pair of atoms, after which the elapsed time can be determined. Practical applications may require $N \gg 1$ in order to achieve a good signal-to-noise ratio.

    In the quantum Zeno effect[11-13], frequent measurements can inhibit transitions into unwanted states and force the system to evolve in the desired subspace of Hilbert space. The Zeno effect has been experimentally demonstrated using $^9Be^+$ ground-state hyperfine levels[14], Bose-Einstein condensates[15], ion traps[16], nuclear magnetic resonance[17], cold atoms[18], cavity QED[19], and large atomic systems[13]. It has also been shown that the Zeno effect is a sufficient resource for the implementation of quantum logic gates[20,21], which could be used as the basis of a quantum computer[21] or quantum repeaters[22]. The Zeno effect can also be used to prepare various nonclassical or entangled states[23-27] and to protect entanglement once it has been generated[28,29]. The anti-Zeno effect, by which repeated measurements increase the rate of transitions, may be useful in quantum heat engines[30].

    Earlier approaches have used entangled states to synchronize distant clocks[31,32]. Our approach is also based on the use of simple entangled states (dark states)[33], but here the goal is to improve the stability

of a single clock rather than synchronize two or more distant clocks. A Zeno-like-effect has recently been shown to mitigate phase diffusion in self-sustaining quantum systems[34-36], which is somewhat related to our technique for inhibiting relative phase drift in atomic clocks. It has also been shown that an atomic clock can be implemented using the entropy reduction produced by weak coupling to the environment and continuous measurements[37].

This paper begins with a discussion of the potential reduction in the bandwidth of an ensemble of atoms using the quantum Zeno effect and the increased precision of an atomic clock that could be achieved in that way. A technique for using the quantum Zeno effect to lock the relative phase of a pair of two-level atoms is then described. The advantages of using three-level or four-level atoms is then discussed, including the ability to directly read out the phase of the atoms after some period of time. The need to extend these methods to larger numbers of atoms is considered.

## Benefit of atomic phase locking

Atomic clocks are typically operated by locking the frequency of an external microwave oscillator to the resonant frequency associated with an atomic transition between two nuclear hyperfine states[38]. Once locked in this way, the external oscillator provides the readout of the clock. The precision of such a clock depends on the bandwidth of the atoms and the measurement interval as described by the Allan deviation[10,39], which is given by

$$\sigma_y(\tau) \approx \frac{\Delta f}{f\sqrt{N}} \sqrt{\frac{T_c}{\tau}}. \qquad (1)$$

Here $\sigma_y(\tau)$ is the standard deviation of the fractional frequency offset $y(t)$ in the output, which is defined by

$$y(t) = \frac{f(t) - f_0}{f_0}. \qquad (2)$$

$T_c$ is the clock cycle time, $\tau$ the total averaging time, and $N$ is the number of independent atoms.

Equation (1) is valid when the statistical noise is much larger than any systematic error. It is based on the assumption that the $k^{th}$ atom in the ensemble has a frequency $f_k$ chosen at random from a normal distribution about a central frequency $f_0$, with a full width at half maximum (FWHM) of $\Delta f$. This corresponds to a standard deviation of $\sigma \equiv \Delta f / 2\sqrt{2 \ln 2}$. The $\sqrt{N}$ term in the denominator of equation (1) corresponds to the fact that the signal-to-noise ratio increases for larger numbers of atoms, since the absorption signal is larger in that case. Equation (1) clearly shows the advantage of reducing the atomic bandwidth $\Delta f$.

Locking an external oscillator to the atomic transitions is roughly equivalent to measuring the spectrum of the atomic transitions and performing a least-square fit to determine the central frequency $f_0$, as illustrated in Figure 1a. For a fixed number of atoms, the peak in the spectrum will be inversely proportional to the bandwidth $\Delta f$, as illustrated in Figure 1b, since the total rate of transitions is fixed. The goal of our approach is to use the quantum Zeno effect to lock the relative phases of all of the atoms, despite the differences between the various values of $f_k$.

After phase locking, the central peak in Figure 1b will correspond to the average $\bar{f}$ of all of the frequencies of the individual atoms, as given by

$$\bar{f} = \frac{1}{N}\sum_{k=1}^{N} f_k. \qquad (3)$$

This results in a standard deviation $\bar{\sigma}$ for $\bar{f}$ that is reduced by a factor of $1/\sqrt{N}$, as is the FWHM $\Delta\bar{f}$, when measured over an ensemble of similarly-prepared systems. The Allan deviation of equation (1) is reduced by the corresponding amount. It should be noted that reducing the bandwidth is in addition to the factor of $1/\sqrt{N}$ that already appears in equation (1), which corresponds to the improved signal-to-noise ratio due to the increased transition rate from $N$ atoms.

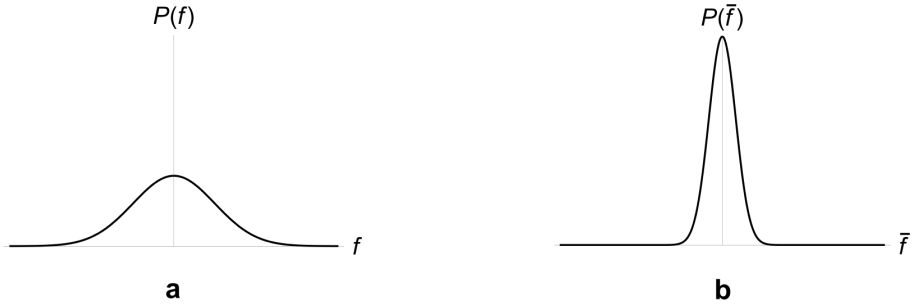

**Figure 1. Reduced bandwidth after phase locking. a,** Initial probability distribution $P(f)$ of the frequencies $f_k$ of $N$ atoms in an atomic clock. **b,** If the frequencies of all of the atoms are locked to their average frequency $\bar{f}$ using the quantum Zeno effect, the width of the probability distribution will be reduced by a factor of $1/\sqrt{N}$ and the height of the peak will be increased accordingly. In this example $N = 9$.

The output of an atomic clock could also be read out by directly measuring the average change in phase of the atoms after they have evolved over a time interval of $\Delta t$. The fact that the atoms have slightly different frequencies will cause a measurement of the average phase to wash out over a period of time. In order to see this, consider a situation in which all of the atoms start out at time $t = 0$ with the same phase. At a subsequent time $t$, the phases of all of the atoms will have evolved independently based on their frequencies $f_k$ as chosen from a normal distribution given by

$$P(f) = \frac{1}{\sqrt{2\pi\sigma^2}} e^{-(f-f_0)^2/2\sigma^2}. \qquad (4)$$

A measurement of the average cosine of the phases will then have the value

$$\cos\phi_{meas.} = \frac{1}{N}\sum_{k=1}^{N}\cos(2\pi f_k t). \qquad (5)$$

The expectation value of this signal measured over a large number of similarly-prepared samples is given by

$$\langle\cos\phi_{meas.}\rangle = \int df\, P(f)\frac{1}{N}\sum_{k=1}^{N}\cos(2\pi f t). \qquad (6)$$

This reduces to

$$\langle \cos \phi_{meas.} \rangle = \int_{-\infty}^{\infty} df\, P(f)\cos(2\pi ft) = e^{-\frac{1}{2}(2\pi t)^2 \sigma^2} \cos(2\pi f_0 t). \quad (7)$$

The exponential factor in equation (7) means that a measurement of the average phase will wash out after a relatively short amount of time, as indicated in Figure 2a.

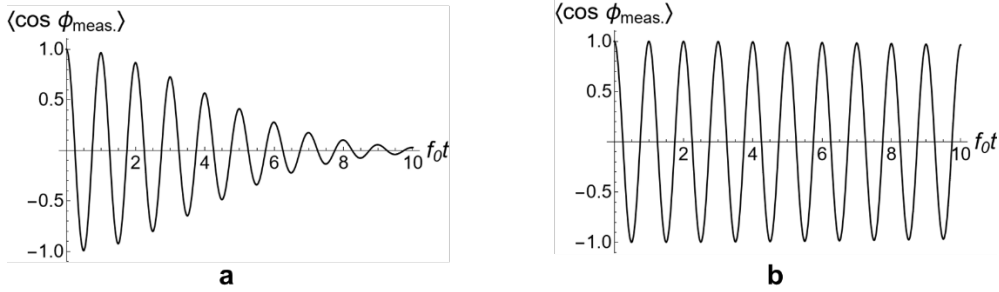

**Figure 2. Average phase of an ensemble of atoms.** The expectation value of the average of the cosine of the phases of an ensemble of atoms is plotted as a function of the time t. **a,** The average cosine of the phase for an ensemble of independent atoms. **b,** The cosine of the phase averaged over an ensemble where the phases of the atoms have all been locked to their average phase $\bar{f}$ using the quantum Zeno effect. For simplicity, the results plotted correspond to $\Delta f / f_0 = 10\%$ and a central frequency of $f_0 = 100 Hz$. It can be seen that the average phase decoheres very rapidly for independent atoms, while locking the phase using the quantum Zeno effect can greatly extend the time interval over which the average phase is coherent and could be measured.

On the other hand, suppose that all of the atoms have been locked onto their average frequency $\bar{f}$ given by equation (3) as a result of the Zeno effect or some other mechanism. $\bar{f}$ is once again a random variable, but with a reduced standard deviation of $\sigma/\sqrt{N}$. Now the cosine of the average phase is simply

$$\cos \phi_{meas.} = \cos(2\pi \bar{f} t), \quad (8)$$

which has an expectation value of

$$\langle \cos \phi_{meas.} \rangle = e^{-\frac{1}{2}(2\pi t)^2 \sigma^2/N} \cos(2\pi f_0 t). \quad (9)$$

In this case the average phase remains coherent over a much longer time interval, as illustrated in Figure 2b. These plots correspond to an arbitrary choice of $N = 100$ atoms with $\Delta f / f_0 = 10\%$.

Figure 2 provides another way to understand the potential benefits of locking the phases of the atoms in an atomic clock. The rest of our analysis will be based on determining the elapsed time by measuring the change in phase using the radiation emitted by the atoms, as will be described below.

## Two-Level Atoms

In this section, we will show that the quantum Zeno effect can be used to lock the relative phase of a pair of two-level atoms. It will be assumed that the two-level atoms interact weakly with a single mode of an optical cavity and that their average frequency $\bar{f}$ is on resonance with the cavity mode, as illustrated in Figure 3. The atoms are initially prepared in a subradiant state $|\psi\rangle$ given by

$$|\psi\rangle = \frac{|EG\rangle - |GE\rangle}{\sqrt{2}} \otimes |0\rangle_{field}, \tag{10}$$

where the state $|EG\rangle$ corresponds to one atom in its excited state and the other atom in its ground state, with a similar notation for the other states of the atoms. The state $|0\rangle_{field}$ represents the vacuum state of the cavity.

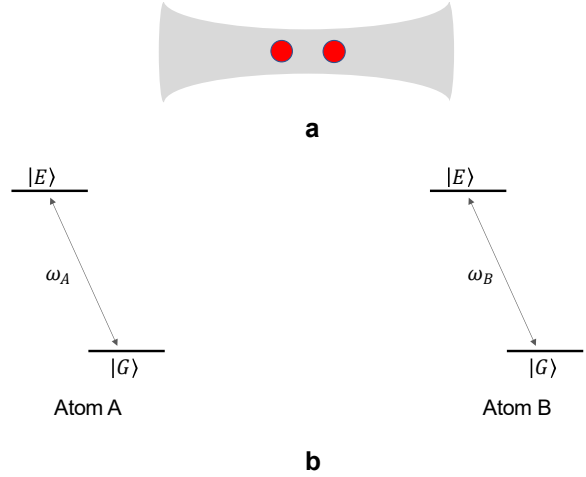

**Figure 3. A pair of two-level atoms in a cavity. a,** A pair of two-level atoms interact weakly with a single mode of an optical cavity. The quantum Zeno effect can be implemented by periodically introducing $n$ photons into the cavity and then measuring the number of photons present after a time interval of $\tau_m$. Alternatively, the atoms could be passed through a cavity containing $n$ photons. **b,** Typical energy level diagrams for the two atoms.

The Hamiltonian for this system in the rotating wave approximation can be written[40] as

$$\hat{H} = \hbar\omega(\hat{n}+1/2) + \hbar\omega_A |E\rangle_A \langle E_A| + \hbar\omega_B |E\rangle_B \langle E_B| \\ + \hbar\frac{\Omega_A}{2}\left[\hat{\sigma}_A^{(+)}\hat{a} + \hat{\sigma}_A^{(-)}\hat{a}^\dagger\right] + \hbar\frac{\Omega_B}{2}\left[\hat{\sigma}_B^{(+)}\hat{a} + \hat{\sigma}_B^{(+)}\hat{a}^\dagger\right]. \tag{11}$$

Here $\omega$ is the angular frequency of the cavity mode while $\omega_A$ and $\omega_B$ are the transition frequencies of the two atoms A and B, which are unequal due to an interaction with the environment. $\Omega_{A(B)}$ are the interaction strengths between the atoms and the photons, and $\hat{\sigma}_{A(B)}^{(\pm)}$ are the atomic raising and lowering operators. In the subsequent analyses, we will assume that the interaction strengths of the two atoms are identical. A pair of atoms in a subradiant state of this kind cannot emit a photon into the cavity mode due to destructive interference between the two probability amplitudes[33,41].

We will now consider the evolution of the initial state of equation (10) over a time interval $\tau$ that is sufficiently short that the effects of the interaction with the cavity modes can be neglected $(\Omega_{A(B)} = 0)$. If the atoms both had the same transition frequency, then the phases of their excited states would evolve at the same rate and the form of the subradiant state would be maintained indefinitely (the minus sign would always hold). Suppose instead that the frequencies of the two atoms are given by $\omega_A = \omega + \delta + \Delta$ and $\omega_B = \omega + \delta - \Delta$. Here $\delta$ is a common deviation from the resonant frequency of the cavity while $2\Delta$ corresponds to a frequency difference between the two atoms due to environmental effects. To first order in $\tau$, the final state of the system is given by

$$|\psi'\rangle = \left\{[1 - i(\omega + \delta)\tau]\left[\frac{|EG\rangle - |GE\rangle}{\sqrt{2}} - i\tau\Delta\frac{|EG\rangle + |GE\rangle}{\sqrt{2}}\right]\right\} \otimes |0\rangle_{field}. \qquad (12)$$

The second term in this equation corresponds to a superradiant state where the rate of photon emission is enhanced by quantum interference effects instead of decreased. It can be seen that any difference in transition frequencies will cause the atoms to gradually develop a superradiant component. The overall phase shift of $-i(\omega + \delta)\tau$ has no physical effects and can be ignored.

The basic idea of our approach is to make frequent measurements to determine whether or not the system has evolved into the superradiant state. If it has not, then the system will collapse back into a pure subradiant state with the original relative phase of 180° as in equation (10). If these measurements are made frequently enough (or continuously), the transition into the superradiant state will be inhibited by the Zeno effect and the relative phase of the two atoms will remain unchanged, with both atoms evolving at the average angular frequency $\bar{\omega} = 2\pi \bar{f}$.

The measurements required for the quantum Zeno effect could be made in several ways. Here we will consider an approach in which the coupling of the atoms to the cavity mode is sufficiently weak that any interaction with the vacuum state (the vacuum Rabi effect) can be neglected. A large interaction with the field can then be produced by temporarily introducing a large number $n$ of photons into the cavity, so that $|0\rangle_{field} \to |n\rangle_{field}$. Alternatively, the atoms could be passed through a small aperture in a cavity that already contains $n$ photons[42,43]. Either way, the coupling to the cavity field can be turned on or off as desired.

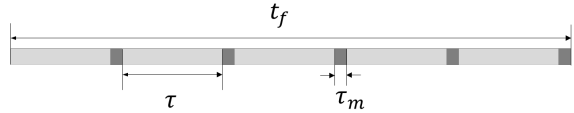

**Figure 4. Relevant time scales in the measurement process.** $\tau$ denotes the time interval between the measurements, during which the system will have evolved a small probability amplitude to be in the superradiant state. A strong interaction between the atoms and $n$ photons in the cavity is applied during the measurement time $\tau_m$, which is chosen to be half a Rabi flop for the superradiant state in equation (12). $t_f$ is the final time at which the output of the clock is read out. We consider the limit of $\tau_m \ll \tau \ll t_f$, where the quantum Zeno effect would be expected to inhibit the growth of the superradiant state and lock the relative phases of the atoms.

The interaction with the field initially containing $n$ photons is maintained for a measurement time $\tau_m$ as illustrated in Figure 4. $\tau_m$ is chosen in such a way that the superradiant term in equation (13) will undergo half a Rabi flop, causing it to absorb or emit one photon[40]. The subradiant term is unaffected by the presence of the photons in the cavity. The initial number of photons $n$ is chosen to be sufficiently large that $\tau_m$ is much shorter than the time required for a significant superradiant amplitude to evolve. This process is repeated until the final time $t_f$ when the output of the clock is read out, as described below. As illustrated in Figure 4, we assume a measurement sequence in which $\tau_m \ll \tau \ll t_f$.

In that limit, we would expect the quantum Zeno effect to inhibit the growth of the superradiant state. Integrating Schrodinger's equation over a time interval of $\tau_m$ using the initial state of equation (12) gives a final state of the form

$$|\psi''\rangle = \frac{|EG\rangle - |GE\rangle}{\sqrt{2}} \otimes |n\rangle_{field}$$
$$-i\tau\Delta \cos\left(\Omega\tau_m\sqrt{n+\frac{1}{2}}\right)\frac{|EG\rangle - |GE\rangle}{\sqrt{2}} \otimes |n\rangle_{field} \tag{13}$$
$$-\frac{\tau\Delta}{\sqrt{2n+1}}\sin\left(\Omega\tau_m\sqrt{n+\frac{1}{2}}\right)\left[\sqrt{n+1}|GG\rangle \otimes |n+1\rangle_{field} + \sqrt{n}|EE\rangle \otimes |n-1\rangle_{field}\right].$$

The second term with the cosine dependence will vanish for half a Rabi flop, while the third term with the sine dependence will reach a maximum. For simplicity, the overall phase factor from equation (12) has been omitted.

It can be seen that the net effect of this measurement process is to change the number of photons by $\pm 1$ if the atoms were in the superradiant state. In principle, the number of atoms could be measured at the end of the time interval $\tau_m$ in order to determine whether or not the superradiant state was present. As a practical matter, no actual measurements are required in the quantum Zeno effect, since the entanglement with the state of the field is sufficient to inhibit the growth of the superradiant state.

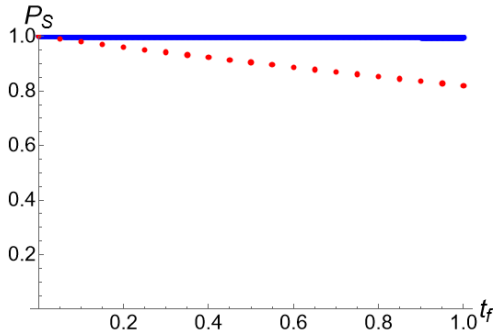

**Figure 5. Plot of the probability $P_S$ that the phase of a pair of two-level atoms will remain locked as a function of time.** The solid line corresponds to relatively frequent measurements with $(\tau + \tau_m) = 0.001$, while the dotted line corresponds to less frequent measurements with $(\tau + \tau_m) = 0.05$. It can be seen that the quantum Zeno effect can lock the relative phase of a pair of two-level atoms in the limit of frequent measurements.

The probability $P_E$ that the system will be found to be in an error state corresponding to one of the last two terms in equation (13) is given by

$$P_E = \Delta^2 \tau^2. \tag{14}$$

The probability $P_S$ of success that the system will remain in the subradiant state at the final time $t_f$ is then given by

$$P_S = (1 - P_E)^{t_f/(\tau+\tau_m)} \approx \exp[-\Delta^2(\tau+\tau_m)t_f]. \qquad (15)$$

$P_S$ is plotted as a function of time in Figure 5 for two different values of $\tau$. With frequent measurements corresponding to $(\tau+\tau_m) = 0.001$, it can be seen that the Zeno effect is very effective in keeping the pair of atoms in the subradiant state with their phases locked to their mean phase. Less frequent measurements corresponding to $(\tau+\tau_m) = 0.05$ are less successful in locking the phase. These results correspond to an arbitrarily chosen value of $\Delta = 2$.

## Three-Level and Four-Level Atoms

In the previous section, we showed that the quantum Zeno effect could be used to lock the relative phase of a pair of two-level atoms. In the limit of frequent measurements, the final state is identical to the initial state aside from an overall phase factor that cannot be measured. As a result, there is no way to read the output of the clock by measuring a time-dependent phase factor. In order to solve this problem, we need a reference state of some kind so that a measurement of the relative phase can be used to estimate the elapsed time as indicated by the clock.

With this in mind, we consider two three-level atoms in a V-configuration as shown in Figure 6. The atoms are assumed to be in an initial state given by

$$|\psi\rangle = [(|E_1 G\rangle - |G E_1\rangle) + (|E_2 G\rangle - |G E_2\rangle)]/2. \qquad (16)$$

Here $|E_1\rangle$ and $|E_2\rangle$ are excited states of the atoms with energies $E_1 \neq E_2$. This state corresponds to a superposition of two subradiant states, and the basic idea is to lock the relative phase of each of the individual terms using the Zeno effect, after which the relative phase between the first and second terms can be used as a readout of the elapsed time.

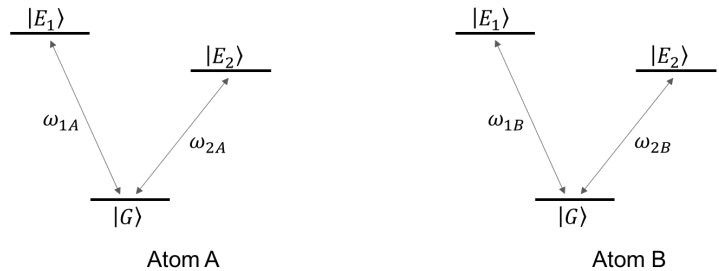

**Figure 6. A pair of three-level atoms in a V-configuration.** A pair of three-level atoms are in a state that corresponds to a superposition of two separate subradiant states, as described in equation (16). This allows the elapsed time to be measured by determining the relative phase of states $|E_1\rangle$ and $|E_2\rangle$.

The Hamiltonian for this system is given by

$$\hat{H} = \hbar\omega_1(\hat{n}_1 + 1/2) + \hbar\omega_2(\hat{n}_2 + 1/2) + \hbar\omega_{1A}|E_1\rangle_A\langle E_1| + \hbar\omega_{2A}|E_2\rangle_A\langle E_2| + \hbar\omega_{1B}|E_1\rangle_B\langle E_1| + \hbar\omega_{2B}|E_2\rangle_B\langle E_2|$$
$$+\hbar\frac{\Omega}{2}\left[\hat{\sigma}_{1A}^{(+)}\hat{a}_1 + \hat{\sigma}_{1A}^{(-)}\hat{a}_1^\dagger\right] + \hbar\frac{\Omega}{2}\left[\hat{\sigma}_{2A}^{(+)}\hat{a}_2 + \hat{\sigma}_{2A}^{(-)}\hat{a}_2^\dagger\right] + \hbar\frac{\Omega}{2}\left[\hat{\sigma}_{1B}^{(+)}\hat{a}_1 + \hat{\sigma}_{1B}^{(-)}\hat{a}_1^\dagger\right] + \hbar\frac{\Omega}{2}\left[\hat{\sigma}_{2B}^{(+)}\hat{a}_2 + \hat{\sigma}_{2B}^{(-)}\hat{a}_2^\dagger\right].$$
(17)

Here $\omega_1$ and $\omega_2$, refer to two cavity modes that are in resonance with the atomic transitions $|E_1\rangle \leftrightarrow |G\rangle$ and $|E_2\rangle \leftrightarrow |G\rangle$, respectively. If the energies of the two atoms are slightly different due to environmental effects, the subradiant states in equation (16) will gradually evolve into superradiant components as in the previous section. The Zeno effect can be used to inhibit the superradiant components as before, where $n$ photons are now introduced into both modes of the cavity.

Unfortunately, it is possible to absorb a photon from the field in a transition such as $|E_1G\rangle|n\rangle_1|n\rangle_2 \leftrightarrow |E_1E_2\rangle|n\rangle_1|n-1\rangle_2$ or $|GE_1\rangle|n\rangle_1|n\rangle_2 \leftrightarrow |E_2E_1\rangle|n-1\rangle_1|n\rangle_2$ even in the subradiant states of Figure 6. This difficulty occurs because both of the excited energy levels interact with a common lower level (the ground state). This problem can be avoided using a pair of four-level atoms as illustrated in Figure 7, where the excited states $|E_1\rangle$ and $|E_2\rangle$ interact with two separate lower-energy states $|G_1\rangle$ and $|G_2\rangle$. These could correspond to two different hyperfine levels of the atomic ground state, for example. The cross-couplings between $|E_1\rangle$ and $|G_2\rangle$ are assumed to be negligibly small, as is the coupling between $|E_2\rangle$ and $|G_1\rangle$.

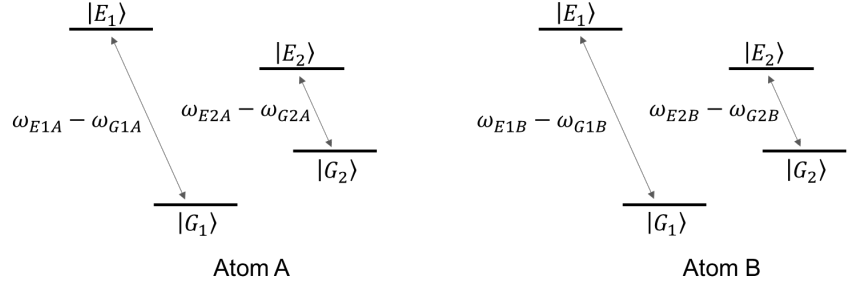

**Figure 7. A four-level atomic system used to avoid transitions from the ground state.** Atomic states $|E_1\rangle$ and $|G_1\rangle$ are coupled by allowed transitions, as are levels $|E_2\rangle$ and $|G_2\rangle$, with negligible coupling between the other states. The quantum Zeno effect is used to lock the relative phase of each pair of states, after which the relative phase between $|E_1\rangle$ and $|E_2\rangle$ can be used to read out the elapsed time from the clock.

The four-level system is prepared in an initial state given by
$$|\psi\rangle = \frac{1}{2}\left[\left(|E_1G_1\rangle - |G_1E_1\rangle\right) + \left(|E_2G_2\rangle - |G_2E_2\rangle\right)\right]|0\rangle_1|0\rangle_2,$$
(18)

with a Hamiltonian given by

$$\begin{aligned}
\hat{H} &= \hbar\omega_1(\hat{n}_1 + 1/2) + \hbar\omega_2(\hat{n}_2 + 1/2) \\
&+ \hbar\omega_{E1A}|E_1\rangle_A\langle E_1| + \hbar\omega_{E2A}|E_2\rangle_A\langle E_2| + \hbar\omega_{G1A}|G_1\rangle_A\langle G_1| + \hbar\omega_{G2A}|G_2\rangle_A\langle G_2| \\
&+ \hbar\omega_{E1B}|E_1\rangle_B\langle E_1| + \hbar\omega_{E2B}|E_2\rangle_B\langle E_2| + \hbar\omega_{G1B}|G_1\rangle_B\langle G_1| + \hbar\omega_{G2B}|G_2\rangle_B\langle G_2| \\
&+ \hbar\frac{\Omega}{2}\left[\hat{\sigma}_{1A}^{(+)}\hat{a}_1 + \hat{\sigma}_{1A}^{(-)}\hat{a}_1^\dagger\right] + \hbar\frac{\Omega}{2}\left[\hat{\sigma}_{2A}^{(+)}\hat{a}_2 + \hat{\sigma}_{2A}^{(-)}\hat{a}_2^\dagger\right] + \hbar\frac{\Omega}{2}\left[\hat{\sigma}_{1B}^{(+)}\hat{a}_1 + \hat{\sigma}_{1B}^{(-)}\hat{a}_1^\dagger\right] + \hbar\frac{\Omega}{2}\left[\hat{\sigma}_{2B}^{(+)}\hat{a}_2 + \hat{\sigma}_{2B}^{(-)}\hat{a}_2^\dagger\right].
\end{aligned} \tag{19}$$

Quantum Zeno measurements can be implemented as before by introducing $n$ photons in each of the two cavity modes that are resonant on the transitions in Figure 7. The photons interact with the atoms for a short time interval $\tau_m \ll \tau$, after which the number of photons in each mode could be measured. (Once again, no actual measurements are required for the Zeno effect.)

The probability $P_E$ that the four-level system will be found to be in an error state corresponding to one of the superradiant states after a time interval of $\tau$ can be shown to be

$$P_E = \left(\frac{\Delta_1^2 + \Delta_2^2}{2}\right)\tau^2. \tag{20}$$

Here $\Delta_k \equiv \left[(E_{kA} - G_{kA}) - (E_{kB} - G_{kB})\right]/2$, which corresponds to the difference in the transition energies of the two atoms. The probability $P_S$ that the system will remain in the subradiant state at the final time $t_f$ is then given by

$$P_S = (1 - P_E)^{t_f/(\tau + \tau_m)} \approx \exp\left[-\left(\frac{\Delta_1^2 + \Delta_2^2}{2}\right)(\tau + \tau_m)t_f\right]. \tag{21}$$

It can be seen that the Zeno effect can inhibit the growth of the superradiant states and maintain a fixed relative phase within each of the states $|E_k\rangle$ and $|G_k\rangle$ as before. The probability of success drops off at the same rates as shown in Figure 5 if we choose $\Delta_1$ and $\Delta_2$ equal to the parameter $\Delta$ in equation (15). Once again, the probability of success can be made arbitrarily high by reducing the value of $\tau$.

## Clock Readout

In our approach, the elapsed time can be read out by measuring the relative phase of the two terms in parentheses in equation (18) at the final time. For large values of $t_f$, multiple oscillations may have occurred, but the output can be used to give a small correction to an external oscillator, for example.

As shown in the previous section, the quantum Zeno effect can be used to eliminate the growth of the superradiant state. In that case, the initial state of equation (18) evolves into

$$|\psi\rangle = \frac{1}{2}\left[\left(|E_1 G_1\rangle - |G_1 E_1\rangle\right) + e^{i\omega_{clock}t_f}\left(|E_2 G_2\rangle - |G_2 E_2\rangle\right)\right] \tag{22}$$

at the final time $t_f$. Here $\omega_{clock} = (\bar{E}_1 + \bar{G}_1 - \bar{E}_2 - \bar{G}_2)/\hbar$, which reflects the phase precession due to the difference in the energy levels. $\bar{E}_k$ and $\bar{G}_k$ are the values of $E_k$ and $G_k$ averaged over the two atoms.

There are several ways to measure the relative phase of these two states. We will consider an approach in which equation (22) is first converted to a superposition of superradiant states, after which the phase of the electromagnetic field emitted by the atoms can be measured. The atoms can be put into a superradiant state by focusing a laser beam on atom B that is slightly off-resonance from the transition between states $|E_1\rangle$ and $|G_1\rangle$. The strength of the interaction can be adjusted to produce a minus sign on state $|E_1\rangle$ for atom B only. A similar procedure can be used to produce a minus sign for state $|E_2\rangle$ of atom B. This converts equation (22) into

$$|\psi'\rangle = \frac{1}{2}\left[\left(|E_1 G_1\rangle + |G_1 E_1\rangle\right) + e^{i\omega_{clock} t_f}\left(|E_2 G_2\rangle + |G_2 E_2\rangle\right)\right]. \tag{23}$$

Now the atoms are in a superposition of superradiant states where the rate of photon emission is enhanced by quantum interference effects instead of being suppressed.

The next step in the readout process is to mix the amplitudes of the two lower ground levels $|G_1\rangle$ and $|G_2\rangle$, which will allow quantum interference in the photon emission process. This can be done using a strong laser beam that is slightly off-resonance from the transition between states $|G_1\rangle$ and $|G_2\rangle$. (If there is no allowed matrix element between these two states, two laser beams and an intermediate state could be used instead.) The strength of the coupling is chosen to give a unitary transition of the form

$$|G_1\rangle \rightarrow \left(|G_1\rangle + |G_2\rangle\right)/\sqrt{2}, \quad |G_2\rangle \rightarrow \left(|G_2\rangle - |G_1\rangle\right)/\sqrt{2}. \tag{24}$$

Inserting this into equation (23) gives

$$|\psi\rangle = \left\{\left[|E_1\rangle\left(|G_1\rangle + |G_2\rangle\right) + \left(|G_1\rangle + |G_2\rangle\right)|E_1\rangle\right] \right. \\ \left. - e^{i\omega_{clock} t_f}\left[|E_2\rangle\left(|G_1\rangle - |G_2\rangle\right) + \left(|G_1\rangle - |G_2\rangle\right)|E_2\rangle\right]\right\}/2\sqrt{2}. \tag{25}$$

We now post-select on the case where the atoms are not in the state $|G_2\rangle$, which could be accomplished by observing the photons emitted in a laser-induced transition to a higher-energy level, as is done in ion-trap quantum computing, for example[44-46]. This reduces equation (25) to

$$|\psi'\rangle = \left[\left(|E_1\rangle|G_1\rangle + |G_1\rangle|E_1\rangle\right) - e^{i\omega_{clock} t_f}\left(|E_2\rangle|G_1\rangle + |G_1\rangle|E_2\rangle\right)\right]/2. \tag{26}$$

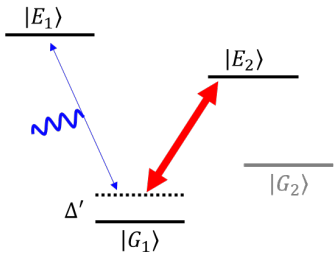

**Figure 8. Measurement of the relative phase of two atomic states.** The excited atomic states $|E_1\rangle$ and $|E_2\rangle$ in equation (26) can be coupled using a strong laser (red arrow) that is detuned from the transition between $|E_2\rangle$ and $|G_1\rangle$. Photons (blue wavy line) can then be emitted in a virtual transition from state $|E_1\rangle$ to $|E_2\rangle$. The phase of the emitted field is shown in Figure 9.

The final step in the readout process is to apply a strong laser beam that is slightly off-resonance from the transition between $|E_2\rangle$ and $|G_1\rangle$, as illustrated in Figure 8. Although the interaction between these two states and the cavity modes was assumed to be negligible, a sufficiently intense laser beam can produce the desired coupling. This allows a virtual transition in which the atoms can emit a photon with a frequency of $\omega_{clock} - \Delta'$ while making a transition from $|E_2\rangle$ to $|E_1\rangle$ [47-49], as shown in Figure 8. Here $\Delta'$ is the detuning of the coupling laser from the ground state.

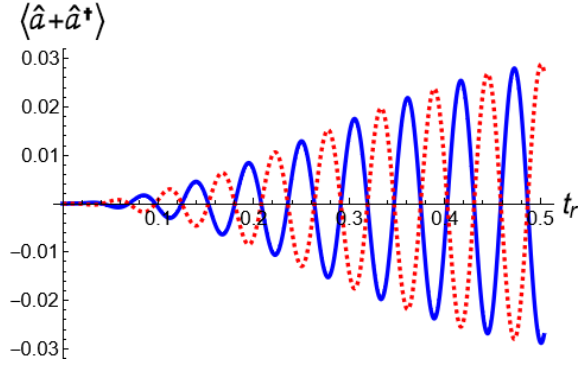

**Figure 9. Mean electric field emitted during the readout of an atomic clock.** The expectation value $\langle \hat{a} + \hat{a}^\dagger \rangle$ of the electric field emitted by the atoms in the state of equation (26) is plotted as a function of the time $t_r$. Here $t_r = 0$ corresponds to the time at which the coupling laser shown in Figure 8 is turned on. The dashed (red) curve corresponds to a relative phase of $\omega_{clock} t_f = 0$, while the solid (blue) curve corresponds to $\omega_{clock} t_f = \pi$. These results correspond to an arbitrary choice of parameters with an interaction strength of $\Omega = 2$, angular frequencies of $\omega_{G_1} = \omega_{G_2} = 0$, $\omega_{E_1} = 1.2 \times 10^2$, $\omega_{E_2} = 1.1 \times 10^2$, a detuning of $\Delta' = 10$, and a coupling laser amplitude of $A = 1$. Dimensionless units have been used for convenience.

The expectation value of the electric field emitted by the atoms in this way is shown in Figure 9 as a function of time, as calculated using first-order perturbation theory. The initial phase of the field is equal to $\omega_{clock} t_f$, which could be directly measured using a homodyne detector, for example. A fit to this data could be used to determine the elapsed time $t_f$ as indicated by the clock.

The accuracy of an atomic clock implemented in this way depends on the uncertainty in $\omega_{clock}$ due to interactions with the environment or other perturbations. The advantage of using the quantum Zeno effect is that locking the phases of the atomic states causes them to evolve at their average frequencies determined by $\bar{E}_k$ and $\bar{G}_k$, which reduces the uncertainty in the clock frequency by a factor of $1/\sqrt{2}$. Practical applications may require that larger numbers of atoms be phase locked in this way in order to achieve an enhancement of $1/\sqrt{N}$.

## Discussion and Conclusions

In this paper, we have considered the possibility of using the quantum Zeno effect to lock the relative phases of an ensemble of $N$ atoms that could be used to implement an atomic clock. This would reduce the effective bandwidth of the ensemble by a factor of $1/\sqrt{N}$ and improve the accuracy of an atomic clock by a corresponding amount.

We began by considering a pair of two-level atoms and showed that the Zeno effect can lock their phases in the limit of frequent measurements. This approach was based on the fact that an initial subradiant state with a small relative phase difference will slowly evolve into a superradiant state with a different relative phase. Frequent observations to determine whether or not the superradiant state has emitted any photons will inhibit the growth of the superradiant state, leaving the atoms in the original subradiant state with a well-defined relative phase. This has the effect of averaging any environmentally-induced phase shifts over the ensemble of atoms.

It was found that using a pair of two-level atoms does not allow a readout of the elapsed time, since there is only an unobservable overall phase in that case. Three-level atoms allow a readout of the elapsed time, but they are susceptible to an error source in which a photon can be absorbed by an atom in the ground state. Both of these difficulties can be resolved by using a pair of four-level atoms, where the elapsed time can be estimated by measuring the relative phase of two different excited states, as illustrated in Figures 8 and 9.

These results show a potential enhancement by a factor of $\sqrt{2}$ in the precision of an atomic clock. Practical applications may require $N \gg 2$ in order to obtain a good signal-to-noise ratio, and we have not been able to generalize this approach to larger values of $N$. In addition, the generation of the required initial state would be difficult for $N \gg 2$, Fock states are required as a resource, and the measurement process is relatively complicated even for $N = 2$. As a result, further research would be required to find a more practical approach. Nevertheless, these results provide an interesting example of the potential use of the quantum Zeno effect.

## Data availability

The Mathematica code used to generate the supporting data is available on request from the corresponding author, S.S.

## References

1　　Major, F. G. *The quantum beat: principles and applications of atomic clocks*. Vol. 2 (Springer, 2007).
2　　Grewal, M. S., Andrews, A. P. & Bartone, C. G. *Global navigation satellite systems, inertial navigation, and integration*. (John Wiley & Sons, 2020).
3　　Blatt, S. *et al.* New Limits on Coupling of Fundamental Constants to Gravity Using $^{87}$Sr Optical Lattice Clocks. *Phys. Rev. Lett.* **100**, 140801 (2008).
4　　Rosenband, T. *et al.* Frequency ratio of Al$^+$ and Hg$^+$ single-ion optical clocks; metrology at the 17th decimal place. *Science* **319**, 1808-1812 (2008).
5　　Martin, M. J. *et al.* A quantum many-body spin system in an optical lattice clock. *Science* **341**, 632-636 (2013).
6　　Bize, S. *et al.* Advances in atomic fountains. *C. R. Phys.* **5**, 829-843 (2004).
7　　Ludlow, A. D., Boyd, M. M., Ye, J., Peik, E. & Schmidt, P. O. Optical atomic clocks. *Rev. Mod. Phys.* **87**, 637 (2015).
8　　Pezze, L., Smerzi, A., Oberthaler, M. K., Schmied, R. & Treutlein, P. Quantum metrology with nonclassical states of atomic ensembles. *Rev. Mod. Phys.* **90**, 035005 (2018).
9　　Pedrozo-Peñafiel, E. *et al.* Entanglement on an optical atomic-clock transition. *Nature* **588**, 414-418 (2020).

10. Allan, D. W. Statistics of atomic frequency standards. *Proc. IEEE* **54**, 221-230 (1966).
11. Misra, B. & Sudarshan, E. C. G. The Zeno's paradox in quantum theory. *J. Math. Phys.* **18**, 756-763 (1977).
12. Facchi, P. & Pascazio, S. Quantum Zeno dynamics: mathematical and physical aspects. *J. Phys. A* **41**, 493001 (2008).
13. Signoles, A. *et al.* Confined quantum Zeno dynamics of a watched atomic arrow. *Nat. Phys.* **10**, 715-719 (2014).
14. Itano, W. M., Heinzen, D. J., Bollinger, J. J. & Wineland, D. J. Quantum zeno effect. *Phys. Rev. A* **41**, 2295 (1990).
15. Do, H. V., Gessner, M., Cataliotti, F. S. & Smerzi, A. Measuring geometric phases with a dynamical quantum Zeno effect in a Bose-Einstein condensate. *Phys. Rev. Res.* **1**, 033028 (2019).
16. Beige, A. Ion-trap quantum computing in the presence of cooling. *Phys. Rev. A* **69**, 012303 (2004).
17. Zheng, W. *et al.* Experimental demonstration of the quantum Zeno effect in NMR with entanglement-based measurements. *Phys. Rev. A* **87**, 032112 (2013).
18. Fischer, M. C., Gutiérrez-Medina, B. & Raizen, M. G. Observation of the quantum Zeno and anti-Zeno effects in an unstable system. *Phys. Rev. Lett.* **87**, 040402 (2001).
19. Bernu, J. *et al.* Freezing coherent field growth in a cavity by the quantum Zeno effect. *Phys. Rev. Lett.* **101**, 180402 (2008).
20. Beige, A., Braun, D., Tregenna, B. & Knight, P. L. Quantum computing using dissipation to remain in a decoherence-free subspace. *Phys. Rev. Lett.* **85**, 1762 (2000).
21. Franson, J. D., Jacobs, B. C. & Pittman, T. B. Quantum computing using single photons and the Zeno effect. *Phys. Rev. A* **70**, 062302 (2004).
22. Bayrakci, V. & Ozaydin, F. Quantum Zeno Repeaters. Preprint at https://arxiv.org/abs/2206.08785 (2022).
23. Wang, X.-B., You, J. Q. & Nori, F. Quantum entanglement via two-qubit quantum Zeno dynamics. *Phys. Rev. A* **77**, 062339 (2008).
24. Raimond, J.-M. *et al.* Quantum Zeno dynamics of a field in a cavity. *Phys. Rev. A* **86**, 032120 (2012).
25. Lin, Y. *et al.* Preparation of entangled states through Hilbert space engineering. *Phys. Rev. Lett.* **117**, 140502 (2016).
26. Magazzù, L., Jaramillo, J. D., Talkner, P. & Hänggi, P. Generation and stabilization of Bell states via repeated projective measurements on a driven ancilla qubit. *Phys. Scr.* **93**, 064001 (2018).
27. Nodurft, I. C., Brewster, R. A., Pittman, T. B. & Franson, J. D. Optical attenuation without absorption. *Phys. Rev. A* **100**, 013850 (2019).
28. Fasihi, M. A. Entanglement preservation in a system of two dipole–dipole interacting two-level atoms coupled with single mode cavity. *Phys. Scr.* **94**, 085104 (2019).
29. Fasihi, M. A., Khanzadeh, M., Hasanzadeh, P. & Asl, S. E. Protecting the entanglement of two interacting atoms in a cavity by quantum Zeno dynamics. *Euro. Phys. J. D* **75**, 1-9 (2021).
30. Mukherjee, V., Kofman, A. G. & Kurizki, G. Anti-Zeno quantum advantage in fast-driven heat machines. *Commun. Phys.* **3**, 1-12 (2020).
31. Jozsa, R., Abrams, D. S., Dowling, J. P. & Williams, C. P. Quantum clock synchronization based on shared prior entanglement. *Phys. Rev. Lett.* **85**, 2010 (2000).
32. Ilo-Okeke, E. O., Tessler, L., Dowling, J. P. & Byrnes, T. Remote quantum clock synchronization without synchronized clocks. *npj Quantum Inf.* **4**, 1-5 (2018).
33. Dicke, R. H. Coherence in spontaneous radiation processes. *Phys. Rev.* **93**, 99 (1954).
34. Li, W., Es'haqi-Sani, N., Zhang, W.-Z. & Vitali, D. Quantum Zeno effect in self-sustaining systems: Suppressing phase diffusion via repeated measurements. *Phys. Rev. A* **103**, 043715 (2021).
35. Lee, T. E. & Sadeghpour, H. R. Quantum synchronization of quantum van der Pol oscillators with trapped ions. *Phys. Rev. Lett.* **111**, 234101 (2013).
36. Rohn, J., Schmidt, K. P. & Genes, C. Phase synchronization in dissipative non-Hermitian coupled quantum systems. Preprint at https://arxiv.org/abs/2111.02201 (2021).

37    Gangat, A. A. & Milburn, G. J. Quantum clocks driven by measurement. Preprint at https://arxiv.org/abs/2109.05390 (2021).
38    Essen, L. & Parry, J. V. L. The caesium resonator as a standard of frequency and time. *Philos. Trans. Royal Soc. A* **250**, 45-69 (1957).
39    Enzer, D. G., Murphy, D. W. & Burt, E. A. Allan Deviation of Atomic Clock Frequency Corrections: A New Diagnostic Tool for Characterizing Clock Disturbances. *IEEE Trans. Ultrason. Ferroelectr. Freq. Control* **68**, 2590-2601 (2021).
40    Shore, B. W. & Knight, P. L. The jaynes-cummings model. *J. Mod. Opt.* **40**, 1195-1238 (1993).
41    Pavolini, D., Crubellier, A., Pillet, P., Cabaret, L. & Liberman, S. Experimental evidence for subradiance. *Phys. Rev. Lett.* **54**, 1917 (1985).
42    Domokos, P., Brune, M., Raimond, J. M. & Haroche, S. Photon-number-state generation with a single two-level atom in a cavity: a proposal. *Euro. Phys. J. D* **1**, 1-4 (1998).
43    Bertet, P. *et al.* Generating and probing a two-photon Fock state with a single atom in a cavity. *Phys. Rev. Lett.* **88**, 143601 (2002).
44    Kielpinski, D., Monroe, C. & Wineland, D. J. Architecture for a large-scale ion-trap quantum computer. *Nature* **417**, 709-711 (2002).
45    Harty, T. P. *et al.* High-fidelity preparation, gates, memory, and readout of a trapped-ion quantum bit. *Phys. Rev. Lett.* **113**, 220501 (2014).
46    Lekitsch, B. *et al.* Blueprint for a microwave trapped ion quantum computer. *Sci. Adv.* **3**, e1601540 (2017).
47    Ashraf, I., Gea-Banacloche, J. & Zubairy, M. S. Theory of the two-photon micromaser: photon statistics. *Phys. Rev. A* **42**, 6704 (1990).
48    Toor, A. H. & Zubairy, M. S. Validity of the effective Hamiltonian in the two-photon atom-field interaction. *Phys. Rev. A* **45**, 4951 (1992).
49    Linskens, A. F., Holleman, I., Dam, N. & Reuss, J. Two-photon Rabi oscillations. *Phys. Rev. A* **54**, 4854 (1996).

# Acknowledgements

We would like to acknowledge many valuable discussions with Todd Pittman, Cory Nunn, and Ian Nodurft. This work was supported in part by the National Science Foundation under grant number PHY-1802472.

# Author contributions

S.S. and J.F. came up with the idea together. S.S. performed the calculations. Both the authors wrote and reviewed the manuscript.

# Additional Information

**Competing interests:** The authors declare that they have no competing interests.